\documentclass[aps,prl,twocolumn,groupedaddress,showpacs]{revtex4}

\usepackage{graphicx}

\begin{document}

\title{The wave nature of biomolecules and fluorofullerenes}
\author{Lucia Hackerm\"uller}
\author{Stefan Uttenthaler}
\author{Klaus Hornberger}
\author{Elisabeth Reiger}
\author{Bj\"orn Brezger}
\altaffiliation{New address: Fakult\"{a}t f\"{u}r Physik,
Universit\"{a}t Konstanz, D-78457 Konstanz}
\author{Anton Zeilinger}
\author{Markus Arndt}
\affiliation{Institut f\"{u}r Experimentalphysik, Universit\"{a}t
Wien, Boltzmanngasse 5, A-1090 Wien} \email[]{contact:
zeilinger.office@univie.ac.at                             }

\date{August 29, 2003}

\begin{abstract}
We demonstrate quantum interference for tetraphenylporphyrin, the
first biomolecule exhibiting wave nature, and for the
fluorofullerene $C_{60}F_{48}$ using a near-field Talbot-Lau
interferometer. For the porphyrins, which are distinguished by
their low symmetry and their abundant occurence in organic
systems, we find the theoretically expected maximal interference
contrast and its expected dependence on the de~Broglie
wavelength. For $C_{60}F_{48}$ the observed fringe visibility is
below the expected value, but the high contrast still provides
good evidence for the quantum character of the observed fringe
pattern. The fluorofullerenes therefore set the new mark in
complexity and mass (1632 amu) for de Broglie wave experiments,
exceeding the previous mass record by a factor of two.
\end{abstract}

\pacs{03.65.-w,03.65.Ta,03.75.-b,39.20.+q}

\maketitle


The wave-particle duality of massive objects is one of the corner
stones of quantum physics. Nonetheless this quantum property is
never observed in our everyday world. The current experiments are
aiming at exploring the limits to which one can still observe the
quantum wave nature of massive objects and to understand the role
of the internal molecular structure and symmetry.

Coherent molecule optics was  already initiated as early as in
1930 when Estermann and Stern confirmed de Broglie's wave
hypothesis~\cite{Broglie1923a} in a diffraction experiment with
He atoms and H$_2$ molecules \cite{Estermann1930a}. In contrast to
the rapidly evolving field of electron and neutron optics, atom
optics became only feasible about twenty years ago and has led
from experiments with thermal atoms to coherent ensembles of
ultra-cold atoms forming Bose-Einstein condensates. {\em Molecule
interferometry} was only taken up again in 1994 with the first
observation of Ramsey-Bord\'e interferences for I$_2$
\cite{Borde1994a} and with the proof of the existence of the
weakly bound He$_2$ in a far-field diffraction experiment
\cite{Schollkopf1994a}. Experiments with alkali dimers in the
far-field \cite{Chapman1995b} and in near-field
\cite{Clauser1997c} interferometers followed. Recent interest in
molecule optics has been stimulated by the quest for
demonstrations of fundamental quantum mechanical effects with
mesoscopic objects \cite{Arndt1999a,Arndt2002a,Bruch2002a}.

\begin{figure}
 \includegraphics[width=1\columnwidth]{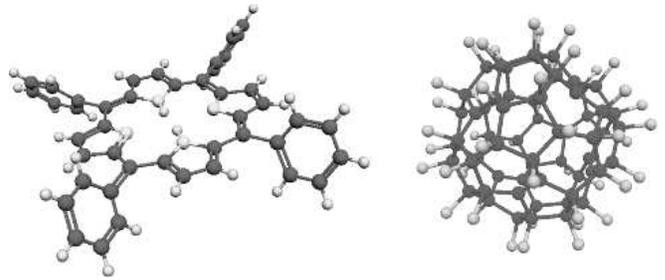}
 \caption[setup]{\label{fig:Figure1} 3D Structure of
 tetraphenylporphyrin (TPP) C$_{44}$H$_{30}$N$_4$  (left) and the
 fluorofullerene C$_{60}$F$_{48}$  (right)\cite{Troyanov2001a}.
 TPP (m=614 amu) is composed of four tilted phenyl rings attached
 to a planar porphyin structure. The fluorofullerene\,(m=1632
 amu) is a deformed C$_{60}$ cage surrounded by a shell of 48
 fluorine atoms. Only an isomer with D$_3$ symmetry is drawn here.
 }
 \end{figure}

In the present letter, we report the first demonstration of the
wave nature of both tetraphenylporphyrin (TPP) and of fluorinated
fullerenes using near-field interference. The porphyrin structure
is at the heart of many complex biomolecules, serving as a color
center for instance in chlorophyll and in hemoglobin. The
fluorofullerene $C_{60}F_{48}$ is the most massive (1632 amu) and
most complex (composed of 108 atoms) molecule for which the
de~Broglie wave-nature has been shown so far (see
fig.\ref{fig:Figure1}).

In order to demonstrate the wave property of a massive object with
a short de Broglie wavelength it is advisable to use a near-field
diffraction scheme.  In particular a Talbot-Lau-interferometer
(TLI, for details see
\cite{Patorski1989a,Clauser1994a,Brezger2002a,Brezger2003a}) is
compact and rugged, has favorable scaling properties,  permits to
work with acceptable grating constants even for large molecules
and allows one to work with an initially uncollimated and
spatially incoherent beam. The basic structure of our
interferometer has been described elsewhere \cite{Brezger2002a}.
The experiment is set up in a vacuum chamber at a base pressure
of $2\times10^{-8}\,$mbar, which is sufficient to avoid molecule
loss or decoherence by residual gas scattering in the
interferometer \cite{Hornberger2003a}. The molecular beam is
created by sublimation in an oven.  TPP was heated to 690 K
corresponding to a vapor pressure of 46 Pa~\cite{Perlovich2000a}.
$C_{60}F_{48}$ was sublimated at 560 K with a vapor pressure of
2.3~Pa~\cite{Boltalina1999a}.

The initial thermal velocity distribution is rather broad (full
width at half maximum $\sim 50$\,\%), both for TPP and for
C$_{60}$F$_{48}$.  We therefore apply a gravitational
$v$-selection scheme: three horizontal slits restrict the beam to
a well-defined free-flight parabola. The first slit is given by
the orifice of the oven ($200\,\mu$m high). The central height
limiter is situated at 138 cm (for TPP) and 126 cm respectively
(for C$_{60}$F$_{48}$) behind the oven. Its opening is set to
$150\,\mu$m. The third horizontal slit ($100\,\mu$m) is
positioned in 10 cm distance from the molecule detector. For
porphyrin we achieve  a velocity resolution between $\Delta v /
v_{m}=30$\,\% (full width at half maximum)  at a mean velocity of
v$_m$=160\,m/s  and $\Delta v / v_{m}=40$\,\% at v$_m$=235\,m/s.
For  $C_{60}F_{48}$ the value is $\Delta v /v_{m} = 20\,\% $ at
v$_m$=105\,m/s. The velocities can be varied by changing the
vertical position of the source and they are measured using a
time-of-flight method.

 The interferometer itself consists of three identical gold gratings
with a grating period of $g = 991.3 \pm 0.3\,$nm, a nominal open
fraction of $f = 0.48 \pm 0.02$ and a thickness of b = 500\,nm
(Heidenhain, Traunreut). The first grating prepares the transverse
coherence of the molecular beam. The second grating is
responsible for the diffraction and interference. The third
grating is used to mask the molecular interference pattern and is
therefore already part of the detection scheme, providing high
spatial resolution. The distance $L$ between the first and the
second grating and that between the second and the third grating
are equalized to within 100 $\mu$m and are of the order of the
Talbot-length $L_T=g^2/\lambda_{\rm{dB}}$ for the most probable
wavelength of the thermal molecular beam. They are set to 0.22 m
for the porphyrins and to 0.38 m  for $C_{60}F_{48}$.  The
alignment and distance of the gratings have been carefully set in
a calibration experiment using $C_{70}$. To record the molecular
interferogram the third grating is scanned across the molecule
beam using a piezo-electric translation stage. The number of
transmitted molecules as a function of the third grating position
is then the measure for the spatial molecular distribution.

The quantum wave nature of the resulting molecule pattern is
reflected in the  strong wavelength dependence of the fringe
visibility.

A previous Talbot-Lau setup had proven successful before
\cite{Brezger2002a} but the detector was limited to the special
case of fullerenes since it employed a thermal laser ionization
scheme which cannot be applied to those molecules which have an
ionization energy exceeding the binding energy. In contrast to
these earlier experiments, all results presented here were
obtained using electron-impact ionization. Although this
ionization mechanism has a much lower efficiency than the optical
scheme it is more universal and can be applied to virtually any
molecule up to a mass of about 2000 amu. The universality of
electron impact ionization is important but it also implies the
necessity of a mass selection stage to prevent the residual gases
in the vacuum chamber from contributing to the signal. In our
experiment this selection is done using a quadrupole mass
spectrometer (ABB Extrel, 2000 amu).

\begin{figure}
 \includegraphics[width=1.00\columnwidth]{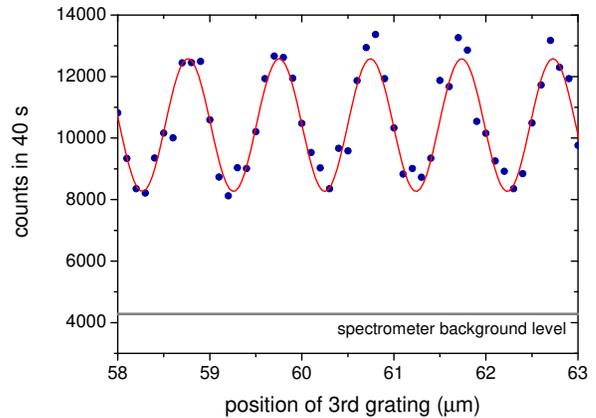}
 \caption[setup]{\label{fig:Figure2}
De Broglie near-field interference fringes of
meso-tetraphenylporphyrin (TPP) resulting from an average over 20
scans at mean velocity of v$_{m}$=160 m/s (dots). Line: Fit with
a sine function. The resulting visibility compares favorably with
the theoretical expectation of $34 \pm 3 \,\%$ . }
 \end{figure}


The first set of experiments reported here were done with
meso-tetraphenylporphyrin (TPP), $C_{44}H_{30}N_{4}$, with a
purity of 98\,\% as purchased from Sigma-Aldrich
(fig.~\ref{fig:Figure1}, left).  It has a diameter of
$\sim$\,20\,\AA, which is about twice the size of C$_{60}$. The
flat geometry of TPP (aspect ratio $\sim\,7$) is rather different
from the highly symmetric structure of the fullerenes which were
used in previous studies. One might imagine two possible effects
of the asymmetric structure which can influence the interference
contrast. First, there could be an
 orientation-dependent coupling between the molecules and the
gratings. Molecules with an anisotropic polarizability will
experience an angle dependent phase shift due to the van der
Waals interaction with the grating walls. This may change the
 interference contrast considerably (see
e.g.~\cite{Brezger2002a}). However, at several hundred Kelvins
the molecules rotate at a frequency of about $10^{13} s^{-1}$
which should lead to an effective averaging of orientation
dependent effects.
 Second, the structure might give rise to
'internal' decoherence. This could be the case if one molecular
axis acted as the hand of a pointer showing finally in different
directions depending on which path the molecule took through the
interferometer. If stray field gradients vary on the scale of the
molecular path separation, as for instance caused by patch fields
in the grating slits, they might encode which-path information in
the internal (e.g. rotational) state. The electric or magnetic
moments required for the coupling to the local fields usually
grow with increasing size and decreasing symmetry of the
molecules.

 To confirm the absence of any
effect of the molecular geometry on the de Broglie interference
it is important to reach experimentally the fringe contrast which
is predicted from the description of the center of mass motion
and the scalar molecular properties alone.

Fig.~\ref{fig:Figure2} now shows the TPP interference fringes
obtained for a mean velocity of v$_{m}$=160 m/s as the third
grating is shifted in steps of 100\,nm across the molecular
density profile. At each step the intensity was recorded over a
period of 2 seconds. This figure is the average of 20 scans. For
our particular grating design quantum mechanics predicts a nearly
sinusoidal molecule intensity as a function of the grating
displacement \cite{Brezger2003a}. We therefore fit a sine
function on top of a mean count rate and a fixed, measured
background to the data. The fringe visibility is then determined
as the ratio of the amplitude to the mean count rate of this fit.
The observed contrast is very close to the theoretically
predicted value of $34 \,\%$, and differs significantly from the
classically expected value of $14\,\%$ (see discussion below).

\begin{figure}
 \includegraphics[width=1.00\columnwidth]{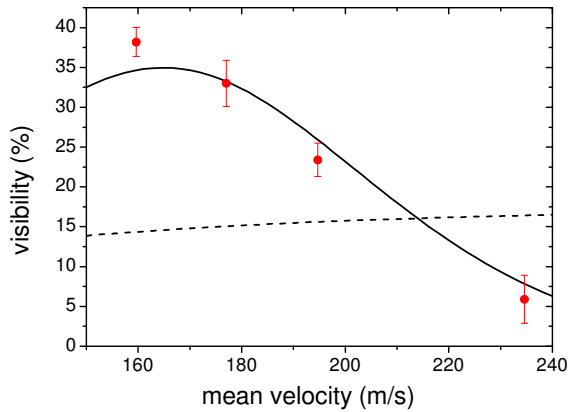}
 \caption[setup]{\label{fig:Figure3}
Velocity dependence of the contrast of the TPP-interference
fringes. The experimental data (full circles) are rather well
described by the quantum mechanical prediction (continuous line),
which is based on wave propagation and takes also into account
the Casimir-Polder interaction between the molecules and the
grating walls. A classical model  - also including the attractive
molecule/grating force - is shown as a dashed line and it fails in
describing the experiment. }
 \end{figure}

In order to further support the interpretation of our data as
being due to the molecular wave-nature we show in
fig.~\ref{fig:Figure3} (full dots)  the wavelength (i.e.
velocity) dependence of the fringe contrast. Four different
velocity distributions were selected from the initial thermal
distribution. For each of them the contrast was determined by
averaging 10 scans as described above. The quantum mechanical
calculation (continuous line) - which includes the molecule/wall
interaction \cite{Brezger2003a} - fits the experimental data
rather well within the experimental errors. The error bars  in
fig.~\ref{fig:Figure3} give the measured statistical standard
deviation within a series of 10 scans \cite{Error}.

However, one might argue that a periodic density pattern behind
grating three could in principle also result from classical
dynamics. Simple shadow images of two gratings are commonly known
as moir\'e fringes. They also depend slightly on the velocity of
the passing object if we take into account the molecule-wall
interaction. But the classically predicted contrast differs
significantly from the measured visibility both in magnitude and
in velocity dependence (fig.\ref{fig:Figure3}, dashed line).

Having proven the perfect contrast for a molecule of low symmetry
it is encouraging to investigate a more complex object.
$C_{60}F_{48}$ is among the heaviest objects that can be
thermally vaporized and detected in our mass spectrometer. It was
purchased from L. Sidorov (State University Moscow) and had a
specified purity of 95\,\%. The fluorofullerenes are present in
several isomers with D$_3$ and S$_6$ symmetry. Typical count
rates were only about 100 counts per second (cps) including a
spectrometer background of 70\,cps with a noise of 30\,\%.  In
addition the background increased up to 130\,cps towards the last
recording. The much reduced signal-to-noise ratio (with respect
to TPP), the extended measuring time and the correspondingly
larger drifts require a very careful data evaluation:

The total intensity is recorded as a function of the position of
the third grating, which was shifted in steps of 40\,nm over a
distance of $3\,\mu$m. In total a series of 68 scans were
recorded over 5 hours. Every second scan was a background scan,
where the source was blocked. Repeated background measurements
were necessary since the dark count rate was changing over time.
The single scans are shifted with respect to each other  due to
thermal drifts of the gratings of up to a few 100 nm between
subsequent scans. Therefore a simple average of all data is not
reasonable.

 The best fitting sine curve
of each single interference pattern was evaluated using a Fourier
transformation and each pattern was shifted to a common origin
according to its phase. Averaging over all recorded scans,
regardless of their individual quality, yields an interference
contrast of 13\,\%, which would still be explicable in classical
terms. However, this includes scans with high background noise
which show no interference at all and which lower the resulting
contrast considerably. In order to eliminate the noisy graphs
without introducing an additional bias by the selection, all
scans were sorted according to their $\chi^2$-value obtained from
the Fourier transformation. The $\chi^2$-value is here a measure
for the distribution of spatial frequencies in the molecular
density pattern. A large $\chi^2$ occurs if there is more than a
single relevant spatial frequency, i.e. it is increasing with
increasing deviation from the expected sine wave shape of the
interferogram.

\begin{figure}
 \includegraphics[width=1.00\columnwidth]{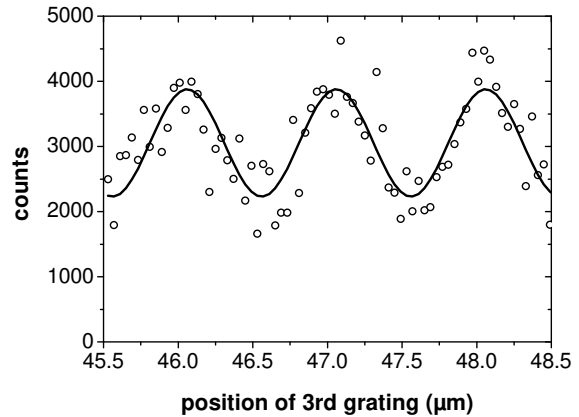}
 \caption[setup]{\label{fig:Figure4}
Quantum interference fringes of $C_{60}F_{48}$. The beam has a
mean velocity of $v_m = 105\,$m/s and a velocity spread (FWHM) of
$\Delta v / v_{m} = 20\,\%$.  To obtain this pattern 14 scans
with the lowest noise were selected and summed after subtracting
the individually measured background (see text). The observed
interference contrast of 27\,\% lies significantly above the
value of 12\,\% expected by a classical model.}
 \end{figure}

It is important to note that this procedure selects recordings
with a clear period, but it explicitly avoids any restrictions on
the particular value of this period. Taking only the fraction $R$
of all scans with the lowest $\chi^2$-value we find an increasing
interference contrast for decreasing $R$. A maximal contrast of
about $27\,\%$ is observed for $R=0.1\dots 0.5$, while single
scans are too noisy to reach this value. We estimate the error of
the visibility for the selected set to $\pm\,3\,\%$. It is also
significant and important that the only high contrast variation
in the molecular density pattern is found at the expected spatial
period of $\sim 1\,\mu$m.
 Figure~\ref{fig:Figure4} shows three periods of the
interference fringes of $C_{60}F_{48}$ selected with $R=0.4$.

It is important to note that this procedure selects recordings
with a clear period but it explicitly avoids any restrictions on
the particular value of this period. Taking only those 40\,\% of
all scans with the lowest $\chi^2$-value we find an interference
contrast of $27\,\%$. The statistical error of the selected set
amounts to $\pm\,3\,\%$. It is also significant and important
that the only observed  high contrast variation in the molecular
density pattern is found at the expected spatial period of $\sim
1\,\mu$m.
 Figure~\ref{fig:Figure4}
shows three periods of the thus selected interference fringes of
$C_{60}F_{48}$.

This high visibility is good evidence for the quantum wave
character of the interference fringes since it lies significantly
above the classically expected contrast of 12\,\%.
 However, the expected quantum mechanical value of $36 \pm 3\, \%$ is still higher
than the experimental contrast. Some information on the origin of
this discrepancy is provided by the fact that pure C$_{70}$,
which can be detected using cw-laser ionization and which does
not suffer from any significant detector noise, exhibits perfect
interference in the same Talbot-Lau setup at a velocity of 200
m/s but it shows a contrast reduction by up to 40\,\% with
respect to the numerical calculations for molecules as slow as
the fluorofullerenes, i.e. with v $\sim 100$ m/s.

We are therefore led to assume that the reason for the difference
between model and experiment is independent of the molecular
species. While we cannot exclude unexpected contributions to the
molecule-wall interaction or more fundamental phenomena, it is
certain that mechanical vibrations of the setup become more
relevant for longer transit times. The strong influence of floor
vibrations was already recognized in earlier fullerene
interferograms and could be drastically reduced by the pneumatic
vibration isolation of the optical table which supports the
experiment. But one has to note that grating motions around 100 nm
are already detrimental to the interference contrast and slow
relative motions can not be excluded with our available vibration
sensors.  Further investigations are required to clarify this
point.

Concluding, we have demonstrated the wave nature of both the
biomolecule TPP and the fluorofullerene $C_{60}F_{48}$ in a
near-field interferometer. The experimental visibilities of the
interference fringes have been compared to a quantum and a
classical model, both taking into account the grating/wall
interaction. For the porphyrin experiment the visibility is in
full agreement with predictions by quantum mechanics and in clear
disagreement with the classical model. This is also the case for
the velocity dependence, i.e. wavelength dependence,  of the
contrast. For $C_{60}F_{48}$ the measured interference contrast
is somewhat below the maximally expected quantum visibility but
it is still significantly above the classical value. This result
provides very good evidence for the wave nature of
fluorofullerenes which are therefore currently the largest and
most complex molecules to show quantum interference.

This work has been supported by the Austrian Science Foundation
(FWF), within the projects START Y177 (M.A.) and SFB F1505, by
the DFG Emmy Noether program (K.H.), as well as by the European
Community under contract HPRN-CT-2000-00125 (E.R.) and
HPMF-CT-2000-00797 (B.B.). We thank O. Boltalina for the
preparation of $C_{60}F_{36}$ for preliminary studies.


\begin{thebibliography}{17}
\bibitem{Error}
The mean visibilities of different scan series  varied due to
uncontrolled external perturbations.
 The additional error is comparable to the statistical one.
\end{thebibliography}

\end{document}